\documentclass[aps,prl,twocolumn,showpacs]{revtex4-1}
\usepackage{graphicx}
\usepackage{amssymb}
\usepackage{amsmath}
\usepackage{appendix}
\usepackage{comment}

\begin{document}

\title{Interfacial Spin and Heat Transfer between Metals and Magnetic Insulators}

\author{Scott A. Bender}
\author{Yaroslav Tserkovnyak}
\affiliation{Department of Physics and Astronomy, University of California, Los Angeles, California 90095, USA}

\begin{abstract}
We study the role of thermal magnons in the spin and heat transport across a normal-metal/insulating-ferromagnet interface, which is beyond an elastic electronic spin transfer. Using an interfacial exchange Hamiltonian, which couples spins of itinerant and localized orbitals, we calculate spin and energy currents for an arbitrary interfacial temperature difference and misalignment of spin accumulation in the normal metal relative to the ferromagnetic order. The magnonic contribution to spin current leads to a temperature-dependent torque on the magnetic order parameter; reciprocally, the coherent precession of the magnetization pumps spin current into the normal metal, the magnitude of which is affected by the presence of thermal magnons.
\end{abstract}



\maketitle

{\em Introduction.}|The excitation of magnetic dynamics by spin-transfer torque \cite{BergerPRB96, *slonczewskiJMMM96,*ralphJMMM08}, and the reciprocal process of spin pumping \cite{tserkovPRL02sp,*tserkovRMP05}, are essential components in the field of metal-based spintronics. Interfacial spin-transfer torque was first realized in conducting magnetic heterostructures, such as spin valves, wherein angular momentum is exchanged with the magnetic order as spin-polarized electrons traverse the structure \cite{tsoiPRL98,*myersSCI99}. Subsequent was the demonstration of the electrical coupling of magnetic insulators interfaced with normal conductors, wherein itinerant-electron spins interact with the magnetic order over atomistically short length scales near the interface. This broadens the ferromagnetic-resonance linewidth \cite{heinrichPRL11,*burrowesAPL12}, allows for spin Hall generation and detection of magnetic dynamics by electrical means \cite{sandwegAPL10,*sandwegPRL11,*hamadehCM14}, and engenders spin Seebeck and Peltier effects that couple magnetic dynamics with heat currents \cite{uchidaNATM10,*bauerNATM12,*flipsePRL14}. Despite this tremendous experimental progress, the basic theoretical problem of the \textit{finite-temperature} transfer of angular momentum across thermodynamically biased normal-metal (N)/ferromagnetic-insulator (F) interfaces remains open.

The out-of-equilibrium magnonic spin transport in an N/F bilayer is well understood in the case when the spin accumulation in N is $collinear$ with the magnetic order parameter in F \cite{benderPRL12,benderPRB14}. In this Letter, we develop a complete description of spin and heat transfer from both the large-angle coherent (i.e., magnetic order) and small-angle incoherent (i.e., magnons) dynamics in F, including the interplay of the two (wherein magnon transport exerts a torque on the magnetic order parameter and vice versa), for arbitrary relative orientations of the N spin accumulation and F magnetization. While the longitudinal spin density injected into F from N is absorbed by the thermal cloud of magnons, the net transverse spin current has to be accommodated by the dynamical reorientation of the ferromagnetic order parameter, i.e., a (temperature-dependent) torque. The strength for both processes can be parameterized by the spin-mixing conductance \cite{brataasPRL00}, which we relate to the quantum-mechanical matrix elements describing elastic and inelastic electron scattering off of the N/F interface.

\begin{figure}[pb]
\includegraphics[width=0.75\linewidth,clip=]{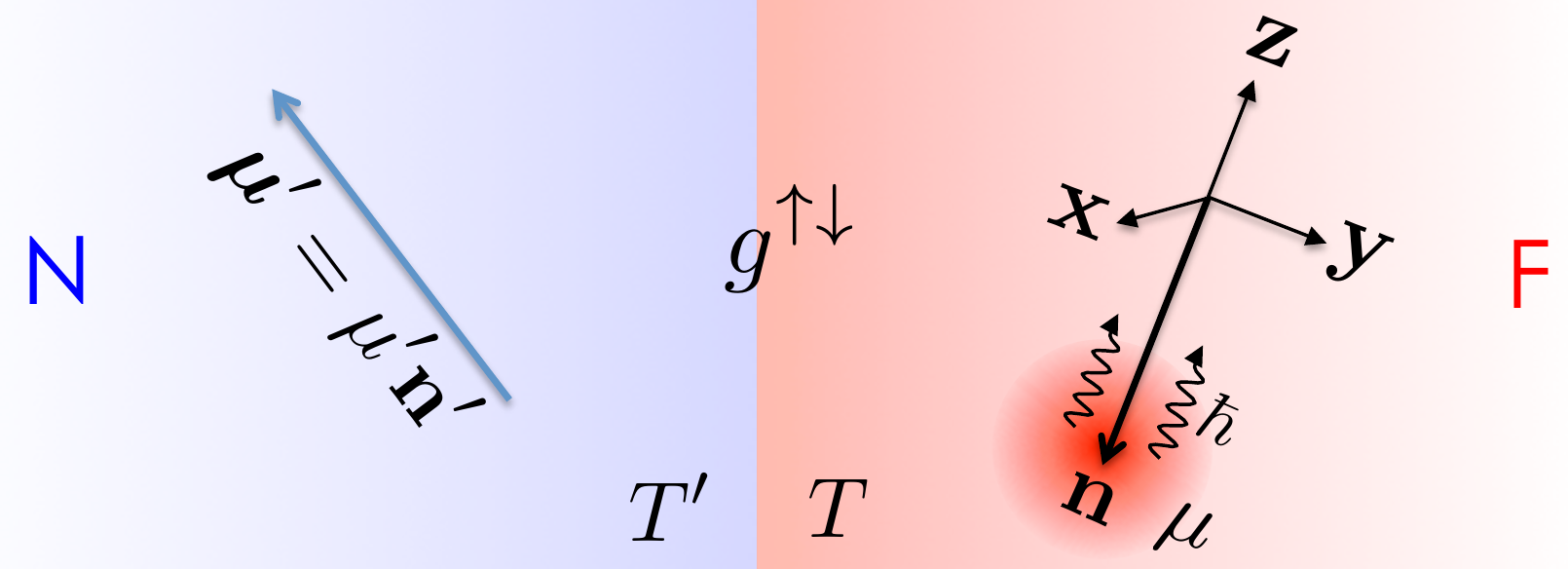}
\caption{Schematic of the N/F junction. $\mathbf{n}$ is the orientation of the ordered spin density in F and $\mathbf{n}'$ is the spin-accumulation direction in N, both near the interface. Itinerant electrons carrying spin $\pm\hbar/2$ along $\mathbf{n}'$ transfer angular momentum via exchange coupling with both the (macroscopic) order parameter $\mathbf{n}$ and magnons in F, the latter each carrying angular momentum $\hbar$ in the $-\mathbf{n}$ ($=\mathbf{z}$) direction and obeying a Bose-Einstein distribution with chemical potential $\mu$. Spin and heat currents across the interface are driven by the out-of-equilibrium spin accumulation $\mu'=\mu_+-\mu_-$ in N, chemical potential $\mu$ in F, and/or an effective interfacial temperature drop $\delta T=T-T'$. The interfacial exchange coupling is quantified by the spin-mixing conductance $g^{\uparrow\downarrow}$ (see text).}
\label{sch}
\end{figure}

{\em Main results.}|We start by summarizing our main results for spin and energy transport across an N/F interface (see Fig.~\ref{sch} for a schematic), where the spins of itinerant electrons in N are exchange coupled to the magnetic moments of F. In N, the out-of-equilibrium spin density (in units of $\hbar$) $\boldsymbol{\rho}$ corresponds to spin accumulation $\boldsymbol{\mu}'\equiv\mu'\mathbf{n}'=2 \boldsymbol{\rho}/D$, where $\mathbf{n}'$ is a unit vector, $D$ is the density of states (per spin and  unit volume), and $\mu'\equiv\mu_+-\mu_-$ is the difference in electrochemical potentials for the electrons up and down along $\mathbf{n}'$ \cite{brataasPRL00}. In the absence of coupling with the ferromagnet, the electronic distribution function, $\langle c^\dagger_{k'\sigma'} c_{k\sigma} \rangle=f_{k\sigma\sigma'} \delta_{kk'}$, can be written as $\hat{f}_k=\sum_{a=\pm}\hat{u}_a f_{ka}$, where $\{\hat{f}_{k}\}_{\sigma\sigma'}=f_{k\sigma\sigma'}$, $\hat{u}_\pm=\left(\hat{1}\pm\mathbf{n}' \cdot \hat{\boldsymbol{\sigma}} \right)/2$ are spin-projection matrices, and $f_{ka}=[e^{\beta'(\epsilon_k-\mu_a)}+1]^{-1}$ is the Fermi-Dirac distribution function with a common temperature $T'\equiv1/\beta'$ (setting $k_B=1$). $\epsilon_k$ is the single-electron energy and $T'$ is assumed to be much smaller than the Fermi energy $\epsilon_F$ so that $D$ can be treated as a constant. 

If the equilibrium spin density (the macroscopic order parameter) in F lies in the direction of the unit vector $\mathbf{n}$, its excitations (magnons) carry $\hbar$ of angular momentum in the $-\mathbf{n}$ direction (neglecting dipolar and spin-orbit interactions, which is possible when the ambient temperature is much larger than the associated energy scales). Built in our treatment is the assumption that magnons maintain an internal thermal equilibrium, which can be achieved, for example, by strong magnon-magnon scattering  (which is enhanced at high temperatures) or by coupling to an external heat sink (e.g., an adjacent copper layer). Thermal magnons in F then follow the Bose-Einstein distribution function: $\langle\hat{a}^\dagger_{q'}\hat{a}_{q}\rangle=n[\beta(\epsilon_q-\mu)]\delta_{qq'}$, where $n(x)=(e^{x}-1)^{-1}$, $\epsilon_q$ is the single-magnon energy, $\mu$ and $T\equiv\beta^{-1}$ are the effective magnon chemical potential and temperature, respectively. This temperature, $T$, understood as the average magnon temperature a correlation length away from the interface, may be different from that of the electrons at the interface, $T'$; the corresponding interfacial thermal bias $\delta T\equiv T-T'$ will affect the flow of spin and heat across the interface.

The relevant variables whose dynamics we wish to study are the vectorial spin density $\boldsymbol{\rho}$ and (scalar-valued) entropy on the electric side and, likewise, vectorial spin density (whose magnitude is determined by the magnonic distribution function and direction by the ordering axis $\mathbf{n}$) and entropy on the magnetic side. The respective thermodynamic forces are $\boldsymbol{\mu}'$ and $T'$ on the N side and $\mu$, $\mathbf{H}\perp\mathbf{n}$, and $T$ on the F side, as will be detailed later. As the central results of this Letter, we calculate, as functions of the temperatures $T$ and $T'$, chemical potentials $\mu$ and $\mu'$, and spin-density orientations $\mathbf{n}$ and $\mathbf{n}'$, the spin and energy currents across the interface.

First, when the magnetic order $\mathbf{n}$ is \textit{static,} we obtain the interfacial spin-current density $\mathbf{i}=-\hbar\dot{\boldsymbol{\rho}}d_N$ out of the normal-metal film of thickness $d_N$, up to first order in $n/s$ (with $n$ as the thermal magnon density and $s$ as the saturation spin density of F in units of $\hbar$):
\begin{equation}
\left.\mathbf{i}\right|_{\dot{\mathbf{n}}=0}=\frac{1}{4\pi}\left(\tilde{g}_i^{\uparrow \downarrow}+\tilde{g}_r^{\uparrow \downarrow}\mathbf{n}\times\right)\boldsymbol{\mu}'\times\mathbf{n}+\tilde{\mathbf{i}}\,,
\label{sc}
\end{equation}
where $\tilde{g}_i^{\uparrow \downarrow}=(1-n/s)g_i^{\uparrow \downarrow}$ and $\tilde{g}_r^{\uparrow \downarrow}=(1-2n/s)g_r^{\uparrow \downarrow}$ \cite{Note1}. $g_r^{\uparrow \downarrow}$ and $g_i^{\uparrow \downarrow}$ here are, respectively, the real and imaginary parts of the $T=0$ spin-mixing conductance per unit area, $g^{\uparrow \downarrow}=g_r^{\uparrow \downarrow}+ig_i^{\uparrow \downarrow}$ \cite{brataasPRL00}. The first term in Eq.~(\ref{sc}) stems from elastic scattering of electrons off of the F macroscopic order, while the last term $\tilde{\mathbf{i}}$ is rooted in thermally-activated electron-magnon scattering at the interface:
\begin{align}
\label{it}
\tilde{\mathbf{i}}=\sum_{a,b=\pm} M_{ab}[&(1-a \mathbf{n}\cdot\mathbf{n}' ) (1+b \mathbf{n}\cdot\mathbf{n}' )\mathbf{n}\\
&+(a/2-b/2+ab \mathbf{n}\cdot\mathbf{n}')\mathbf{n}\times\mathbf{n}' \times \mathbf{n}]\,,\nonumber
\end{align}
with
\begin{align}
M_{ab}=&\frac{g^{\uparrow\downarrow}_r }{4\pi s}\int_{0}^\infty d\epsilon g( \epsilon ) \left( \epsilon+\hbar \Omega -\mu_{ab} \right)\nonumber\\
&\times \left\{ n\left[\beta (\epsilon-\mu^*) \right] -n\left[ \beta' (\epsilon+\hbar \Omega-\mu_{ab}) \right] \right\}\nonumber\, .
\end{align}
Here, $\mu_{ab}\equiv\mu_a-\mu_b$, $g\left(\epsilon \right)$ is the magnon density of states (per unit volume), $\hbar \Omega$ is the magnon gap (so that each magnon carries  energy $E=\epsilon+\hbar \Omega$), and $\mu^*\equiv\mu-\hbar \Omega$. We are supposing that the structure of the thermal magnons is dominated by exchange interactions, such that they are circularly polarized and carry a well-defined spin. Regarding the energy-current density $\dot{q}=-(d/dt)\sum_{k} \epsilon_ k \langle \hat{a}^\dagger_k \hat{a}_k \rangle/A$ out of N through the interfacial area $A$, elastic scattering does not contribute, while inelastic scattering yields:
\begin{equation}
\dot{q}=-\sum_{a,b=\pm}N_{ab} \left(1-a\mathbf{n}\cdot\mathbf{n}' \right) \left(1+b \mathbf{n}\cdot\mathbf{n}' \right) \, ,
\label{tes}
\end{equation}
where
\begin{align}
N_{ab}=&\frac{ g_r^{ \uparrow \downarrow}}{4\pi\hbar s} \int_0^\infty d\epsilon g\left( \epsilon \right) \left( \epsilon+\hbar \Omega \right) \left( \epsilon+\hbar \Omega -\mu_{ab} \right)\nonumber\\
&\times  \left\{n\left[\beta (\epsilon-\mu^*) \right]-n\left[ \beta' (\epsilon+\hbar \Omega-\mu_{ab}) \right]\right\}\,.\nonumber
\end{align}

Second, in order to furthermore include order-parameter \textit{dynamics} $\mathbf{n}(t)$, we specialize the above results to linear response, thus allowing us to utilize the Onsager reciprocity \cite{onsagerPR31} (see discussion below). Our final expressions (assuming weak thermodynamic biases and slow order-parameter dynamics) for the total spin and heat currents (into F) at the interface are:
\begin{align}
\label{Itot}
\mathbf{i}=&\frac{1}{4\pi}\left(\tilde{g}_i^{\uparrow\downarrow}+\tilde{\mathfrak{g}}_r^{\uparrow\downarrow}\mathbf{n}\times\right)(\boldsymbol{\mu}'\times\mathbf{n}-\hbar\dot{\mathbf{n}})\\
&+\left[g \left(\mu+\mathbf{n}\cdot \boldsymbol{\mu}' \right)+\mathcal{S}\delta T\right]\mathbf{n}\,,\nonumber\\
\dot{q}=&-\kappa \delta T-\Pi \left(\mu+\mathbf{n}\cdot \boldsymbol{\mu}' \right)\,.
\label{hclr}
\end{align}
Here,
\begin{equation}
\tilde{\mathfrak{g}}_{r}^{\uparrow \downarrow}=\tilde{g}_r^{\uparrow \downarrow}+ 4\pi  \sum_{a,b=\pm}\partial_{\mu'}M_{ab}\left(a/2-b/2+ab \mathbf{n}\cdot\mathbf{n}' \right)\nonumber
\end{equation}
is the total effective (real part of the) spin-mixing conductance, $g=4\partial_\mu M_{++}$ and $\mathcal{S}=4\partial_T M_{++}$ are respectively the spin conductance and spin Seebeck coefficient, $\kappa=4\partial_T N_{++}$ and $\Pi =4\partial_\mu N_{++}=T\mathcal{S}/\hbar$ are the (magnonic) thermal conductance and spin Peltier coefficient. All these transport coefficients are evaluated in equilibrium and pertain to the interface. $g$, $\kappa$, $\mathcal{S}$, and $\Pi$ are all thermally activated, while $\tilde{\mathfrak{g}}_r^{\uparrow \downarrow}$ and $\tilde{g}_i^{\uparrow\downarrow}$ reduce to the real and imaginary components of the familiar \cite{brataasPRL00,tserkovPRL02sp} zero-temperature spin-mixing conductance $g^{\uparrow\downarrow}$ at $T=0$ and acquire thermal corrections that scale as $\sim (T/T_c)^{3/2}$ when $T\gg \hbar \Omega$ (assuming quadratic magnon dispersion), where $T_c$ is the Curie temperature.

The transverse (i.e., $\perp\mathbf{n}$) component of the spin current \eqref{Itot} exerts a torque on the magnetic order parameter, which enters on the right-hand side of the generalized Landau-Lifshitz equation:
\begin{equation}
(1+\alpha \mathbf{n}\times)\hbar \dot{\mathbf{n}}+\mathbf{n}\times \mathbf{H}=(\alpha'_i+\alpha'_r \mathbf{n}\times)(\boldsymbol{\mu}'\times \mathbf{n}-\hbar \dot{\mathbf{n}})\,,\nonumber
\end{equation}
where $\alpha$ is the bulk Gilbert damping, $\mathbf{H}$ is the effective magnetic field (in appropriate units), $\alpha'_i=\tilde{g}_i^{\uparrow \downarrow}/4\pi \tilde{s}d_F$, $\alpha'_r=\tilde{\mathfrak{g}}_{r}^{\uparrow \downarrow}/4\pi \tilde{s}d_F $, $\tilde{s}=s-n$, and $d_F$ is the ferromagnet's thickness. (Note that the torque depends on $\boldsymbol{\mu}'$ and Onsager-reciprocal spin pumping $\propto\dot{\mathbf{n}}$ but not on $\mu$ or $\delta T$.) The longitudinal (i.e, $\parallel\mathbf{n}$) spin current, on the other hand, is accommodated by the magnon flux into the ferromagnet, $i_m=\dot{n}d_F=-\mathbf{n}\cdot \mathbf{i}/\hbar$, driven by the thermodynamic forces $\mu$, $\boldsymbol{\mu}'$, and $\delta T$:
\begin{equation}
\hbar i_m=-g\left(\mu+\mathbf{n}\cdot \boldsymbol{\mu}' \right)-\mathcal{S}\delta T\,,\nonumber
\end{equation}
which does not depend on the precession of $\mathbf{n}$.

{\em Interfacial coupling.}|As an effective model for the coupling between the spin degrees of freedom of N interfaced with F, we take the exchange  Hamiltonian:
\begin{equation}
\hat{\mathcal{H}}=-J\int d^2\mathbf{r}\,\boldsymbol{\hat{\rho}}\left(\mathbf{r}\right)\cdot\mathbf{\hat{s}}\left(\mathbf{r}\right)\,,
\label{H0}
\end{equation}
where integration is performed over the interfacial area. Here, $\boldsymbol{\hat{\rho}}(\mathbf{x})=\sum_{\sigma\sigma'}\hat{\psi}_{\sigma}^{\dagger}(\mathbf{x})\boldsymbol{\sigma}_{\sigma\sigma'}\hat{\psi}_{\sigma'}(\mathbf{x})/2$ is the N spin density (with $\hat{\psi}_{\sigma}$ being spin-$\sigma$ itinerant electron field operators and $\boldsymbol{\sigma}$ a vector of Pauli matrices) and $\mathbf{\hat{s}}$ is the F spin density associated with localized electron orbitals, both expressed in units of $\hbar$. We will take $-\mathbf{n}$ to be the spin-quantization axis for the electrons in N, such that an itinerant electron with $\sigma={\uparrow}$ (${\downarrow}$) carries an angular momentum of $\hbar/2$ in the $\mp\mathbf{n}$ direction.  Expanding $\hat{\psi}_\sigma(\mathbf{x})=\sum_{k}\psi_k(\mathbf{x})\hat{c}_{k\sigma}$ in terms of the electron annihilation operators $\hat{c}_{k\sigma}$ within an orthonormal basis $\psi_k(\mathbf {x})$ labeled by orbital quantum numbers $k$ (corresponding to spin-degenerate energy eigenstates in the absence of magnetic coupling, $J=0$), we write
\begin{equation}
\boldsymbol{\hat{\rho}}(\mathbf{x})=\frac{1}{2}\sum_{\sigma\sigma' kk'}\psi_{k}^*(\mathbf{x})\psi_{k' }(\mathbf{x})\hat{c}_{k \sigma}^\dagger \boldsymbol{\sigma}_{\sigma\sigma'}\hat{c}_{k' \sigma'}\,.\nonumber
\end{equation}

Orienting a spin-space Cartesian coordinate system for the $z$ axis to point in the $-\mathbf{n}$ direction, we write, via the Holstein-Primakoff transformation \cite{holsteinPR40}, the F spin density:
\begin{equation}
\hat{s}_{z}(\mathbf{x})=\hat{\phi}^{\dagger}(\mathbf{x})\hat{\phi}(\mathbf{x})-s\,,\,\,\,\hat{s}_-(\mathbf{x})=\sqrt{2s-\hat{\phi}^{\dagger}(\mathbf{x})\hat{\phi}(\mathbf{x})}\hat{\phi}(\mathbf{x})\,,\nonumber
\end{equation}
where $\hat{s}_\pm\equiv \hat{s}_{x}\pm i\hat{s}_{y}$ and $\hat{\phi} \left(\mathbf{x} \right)$ is the magnon field operator obeying the bosonic commutation relation $[\hat{\phi} \left(\mathbf{x} \right),\hat{\phi}^\dagger \left(\mathbf{x'} \right)]=\delta \left(\mathbf{x}-\mathbf{x}' \right)$. Expressing $\hat{\phi} \left(\mathbf{x} \right)=\sum_q \phi_q\left( {\mathbf{x}} \right) \hat{a}_q$ in the orthonormal spin-wave basis $\phi_q\left( {\mathbf{x}} \right)$ ($\hat{a}_q$ being the magnon annihilation operators) and inserting the above spin densities into Eq.~(\ref{H0}), we obtain to second order in $\hat{a}_q$:
\begin{align}
\hat{\mathcal{H}}\approx&\sum_{kk'\sigma}U_{kk'\sigma}\hat{c}_{k \sigma}^{\dagger}\hat{c}_{k'\sigma} (1-\hat{n}/s) \nonumber\\
&+\Big(\sum_{k k'q}V_{kk'q}\hat{c}_{k \uparrow}^{\dagger}\hat{c}_{k'\downarrow}\hat{a}_{q}+{\rm H.c.}\Big)\, ,
\label{H1}
\end{align}
where $\hat{n}=\sum_{q}\hat{a}_q^{\dagger}\hat{a}_q/Ad_F$ is the magnon-density operator.

The first term in Eq.~(\ref{H1}) has matrix elements
\begin{equation}
U_{kk'\uparrow}\equiv J\frac{s}{2}\int d^2\mathbf{r}\,\psi_{k}^{*}\left(\mathbf{r}\right)\psi_{k'}\left(\mathbf{r}\right)=-U_{kk'\downarrow}\nonumber
\end{equation}
and describes the elastic scattering of electrons off the static magnetization of F. When the spin of an incoming electron in N is collinear with $\mathbf{n}$, scattering by $U_{kk'\sigma}$ mixes orbital states while preserving the spin, such that no torque is exerted on F. In general, we can supply electrons that are polarized along a different axis $\mathbf{n}'$, as sketched in Fig.~\ref{sch}. Rewriting the first term in Eq.~\eqref{H1} in the corresponding $\pm$ basis, we would obtain the spin-flip terms $\propto\hat{c}^\dagger_{k_+} \hat{c}_{k'_-}$, which result in a spin angular-momentum transfer to F. The second term in Eq.~(\ref{H1}) has matrix elements
\begin{equation}
V_{kk'q}\equiv-J\sqrt{\frac{s}{2}}\int d^2\mathbf{r}\,\psi_{\mathbf{k}}^{*}\left(\mathbf{r}\right)\psi_{\mathbf{k}'}\left(\mathbf{r}\right)\phi_{\mathbf{q}}\left(\mathbf{r}\right)\,,\nonumber
\end{equation}
and, along with its conjugate, describe inelastic spin-flip scattering processes wherein an up-electron/down-hole pair is created (annihilated) in N, along the $z$ axis, destroying (creating) a magnon in F. As we show below, in contrast to elastic scattering processes, such inelastic spin flips generate a temperature-dependent spin current with a component along $\mathbf{n}$.

Having established the equilibrium states of magnons in F and electrons in N (held at different temperatures, $T$ and $T'$, and spin chemical potentials, $\mu$ and $\mu'$) when $J=0$, we treat the transport perturbatively for a finite $J$. To this end, we utilize the Kubo formula to calculate the spin current $\mathbf{i}$, up to second order in exchange $J$, yielding an expression in the form of Eq.~(\ref{sc}). The first term in Eq.~({\ref{sc}}) arises from elastic scattering $U_{kk'\sigma}$, which governs coefficients $g_i^{\uparrow \downarrow}=D U$ and $g_r^{\uparrow \downarrow}=D^2 \left| U' \right|^2$, where
\begin{align}
U\equiv&\frac{2\pi}{AD} \sum_k \delta(\epsilon_F-\epsilon_k) (U_{kk\uparrow}-U_{kk\downarrow})\,,\nonumber\\
\left| U'\right|^2\equiv &\frac{\pi^2}{2AD^2}\sum_{kk'}\delta(\epsilon_F-\epsilon_{k})\delta(\epsilon_F-\epsilon_{k'})\nonumber\\
&\times \left[\left|U_{kk'\uparrow} \right|^2+\left|U_{kk'\downarrow} \right|^2-2\mathrm{Re}\left(U_{kk'\uparrow}U_{kk'\downarrow}^*\right) \right] \,.
\nonumber
\end{align}
Thus the reactive ($g_i^{\uparrow\downarrow}$) and dissipative ($g_r^{\uparrow\downarrow}$) spin currents arising from elastic scattering depend on the orientations of the N and F spin densities but not on thermal bias. From the form of this spin current [i.e., the first term in Eq.~\eqref{sc}], which survives a nonperturbative scattering-matrix treatment \cite{brataasPRL00}, we identify $g_r^{\uparrow\downarrow}$ and $g_i^{\uparrow\downarrow}$ as the real and imaginary parts of the spin-mixing conductance.

In contrast, the magnonic contribution $\tilde{\mathbf{i}}$ to spin current, which arises from inelastic processes $V_{kk'q}$, is additionally dependent on the magnon distribution function in F and temperature in N. To calculate it, let us start by considering the spin current associated with a single mode $q$ at energy $\epsilon_q\ll\epsilon_F$ that is macroscopically occupied with $n_qAd_F\gg 1$ magnons. In this case, using the second line in Eq.~(\ref{H1}), we obtain via the Kubo formula:
\begin{equation}
\tilde{\mathbf{i}}_q=n_q  \left| V_q \right|^2 D^2\left[\mathbf{n}\times \boldsymbol{\mu}'\times\mathbf{n}+2\mathbf{n} (\mathbf{n} \cdot \boldsymbol{\mu}'+ \hbar \Omega)\right]\,,
\label{csc}
\end{equation}
 where 
\begin{equation}
\left| V_q  \right|^2\equiv\frac{\pi d_F}{D^2}\sum_{kk'} \left|V_{kk'q}\right|^2\delta \left(\epsilon_F-\epsilon_k \right) \delta \left(\epsilon_F-\epsilon_{k'} \right)\,.\nonumber
\end{equation}
The macroscopic occupation of the $q=0$ mode is, in essence, a precessing macrospin, which may be excited at zero temperature while all of the thermal modes are frozen out. The spin current \eqref{csc} (with $q=0$) into F may then be compared with the standard expression \cite{tserkovPRL02sp} for spin current $\mathbf{i}_0(t)$ produced by a single classical macrospin pointing in the direction $\mathbf{n}_0(t)$:
\begin{equation}
\mathbf{i}_0\left(t\right)=\frac{1}{4\pi}\left(g_i^{\uparrow \downarrow}+g_r^{\uparrow \downarrow}\mathbf{n}_0\times\right)\left(\boldsymbol{\mu}'\times\mathbf{n}_0-\hbar\dot{\mathbf{n}}_0\right)\,.\nonumber
\end{equation}
Suppose $\mathbf{n}_0(t)$ precesses at a small angle $\theta$ circularly around $\mathbf{n}$ at a frequency $\Omega$. Identifying $n_0=s(1-\cos\theta)\approx s\theta^2/2$, we get for the cycle-averaged spin current (for a constant $\boldsymbol{\mu}'$):
\begin{align}
\langle\mathbf{i}_0\rangle=&\frac{1}{4\pi}\left(g_i^{\uparrow \downarrow}+g_r^{\uparrow \downarrow}\mathbf{n}\times\right)\boldsymbol{\mu}'\times\mathbf{n}-\frac{n_0}{4\pi s}\Big\{g_i^{\uparrow \downarrow}\boldsymbol{\mu}'\times\mathbf{n}\nonumber\\
&+g_r^{\uparrow \downarrow}\left[\mathbf{n}\times\boldsymbol{\mu}'\times\mathbf{n}-2\mathbf{n}(\mathbf{n}\cdot\boldsymbol{\mu}'+\hbar\Omega)\right]\Big\}\,,\nonumber
\end{align}
to first order in $n_0/s$. This classical result is matched with our quantum-mechanical calculation, Eqs.~(\ref{sc}) and~(\ref{csc}), provided we identify:
\begin{equation}
 4\pi s|V_0|^{2}D^2=g^{\uparrow\downarrow}_r\,.
\label{gr}
\end{equation}
Crucially, this remarkable identification is only possible once the matrix elements $U_{kk'\sigma}$ are properly related to the spin-mixing conductance and the $n/s$ corrections are included in $\tilde{g}^{\uparrow\downarrow}$, as described above. 

For thermal magnons with finite wave numbers (that are still much smaller than the Fermi momentum of electrons) normal to the interface \cite{hoffmanPRB13,*kapelrudPRL13}, $|V_q|^2=2|V_0|^2$, because of the Neumann (exchange) boundary conditions at the F film boundaries. Having thus related $|V_q|^2$ to the real part of the spin-mixing conductance, according to Eq.~\eqref{gr}, we finally calculate the magnonic spin current using the second line of Eq.~(\ref{H1}), with the Bose-Einstein distribution for magnons instead of the macroscopic occupation. The resultant expression for the thermal spin and heat currents are given by Eqs.~(\ref{it}) and~(\ref{tes}), respectively. We conclude that the spin-mixing conductance $g^{\uparrow\downarrow}$ captures all the relevant, both elastic and inelastic, matrix elements that govern interfacial spin transport.

{\em Nonequilibrium thermodynamics.}|Supposing that the internal relaxation of thermal magnons is sufficiently fast in comparison with the resonant precessional dynamics $\Omega$ of the macroscopic order parameter $\mathbf{n}$ \cite{benderPRB14}, an instantaneous state of the magnet can be described by three variables: the spin order $\mathbf{n}$, magnon density $n$, and entropy $S_F$. The normal layer is characterized by its spin density $\boldsymbol{\rho}$ and entropy $S_N$. The most natural thermodynamic potential for our purposes is the total internal energy $U(\mathbf{n},n,\boldsymbol{\rho};S_F,S_N)$ of the N/F bilayer as a function of these variables.  These parameters (when conveniently normalized) form the following conjugate pairs with their respective thermodynamic forces: $(Ad_F s \mathbf{n} , \mathbf{H})$, $(Ad_F n, \mu)$, $(Ad_N \boldsymbol{\rho}$, $\boldsymbol{\mu}')$, $(S_F, T)$, and $(S_N,T')$. We now consider the structure of the Onsager-reciprocal relaxation of our system when perturbed away from the equilibrium. Since for a closed system, the total entropy $S=S_F+S_N={\rm const}$, at linear response, only the entropic flow $\delta S=(S_F-S_N)/2$ is relevant, whose thermodynamic conjugate is $\delta T$.

We start by deriving the spin current \eqref{sc} in the absence of the order-parameter dynamics, i.e., $\dot{\mathbf{n}}=0$, which is then entered in the equations of motion for spin densities:
\begin{align}
\label{drho}
\hbar\dot{\boldsymbol{\rho}}d_N&=-\mathbf{i}+{\rm Bloch~relaxation}\,,\\
\hbar\left[(s-n)\dot{\mathbf{n}}-\dot{n}\mathbf{n}\right]d_F&=\mathbf{i}+{\rm LLGB~dynamics}\,,
\label{dn}
\end{align}
where ``Bloch relaxation" stands for the possible spin relaxation inside N and, similarly, ``LLGB dynamics" for the subsequent  Landau-Lifshitz-Gilbert precession of the order parameter along with a Bloch relaxation of magnons. Equations \eqref{drho}, \eqref{dn} could, furthermore, serve as boundary conditions for subsequent spin diffusion carried by electrons and/or magnons away from the N/F interface. According to the Onsager principle, $\boldsymbol{\mu}'$ thus affecting magnetic dynamics $\dot{\mathbf{n}}$ [through the spin current \eqref{sc} on the right-hand side of Eq.~\eqref{dn}] implies that $\mathbf{H}$ must similarly enter in the equation for $\dot{\boldsymbol{\rho}}$. As can be shown \cite{halsEPL10} by straightforward manipulations of Eqs.~\eqref{drho}, \eqref{dn}, this is accomplished by the substitution $\boldsymbol{\mu}'\to\boldsymbol{\mu}'-\hbar\mathbf{n}\times\dot{\mathbf{n}}$ in Eq.~\eqref{sc}, leading finally to Eq.~\eqref{Itot}.

Regarding the longitudinal spin current that is carried by magnons [second line in Eq.~\eqref{Itot}], its Onsager reciprocity with the heat flux \eqref{hclr} is guaranteed by the equivalence between the spin Peltier and Seebeck coefficients, $\Pi=T\mathcal{S}/\hbar$, which arises naturally within our Kubo calculation and is analogous to the so-called second Thompson relation in thermoelectrics. (We remark here that $A\dot{q}=\delta\dot{S}T$, within the linear response.) Note there is no linear response in $\dot{\mathbf{n}}$ to $\mu$ or $\delta T$, nor (reciprocally) is there a linear response in the magnon and heat currents to the precessional order-parameter dynamics, within our exchange approximation.

{\em Outlook.}|Magnon-induced torques and spin currents may manifest in a variety of F/N heterostructures. In general, our expressions for spin and energy currents serve as boundary conditions which must be complemented by the appropriate bulk transport theory for both electrons and magnons.  For example, in heterostructures utilizing spin Hall effect in order to convert between spin and charge currents on the normal-metal side, the temperature-dependent spin currents flowing through the structure in response to a thermoelectric bias (as is, for example, the case in the conventional spin Seebeck effect), as well as the temperature-dependent spin Hall resistance, could be obtained by self-consistently solving the spin Hall diffusion equations in conjunction with Eqs.~(\ref{drho}) and (\ref{dn}) employed at the boundaries. Thermal spin torques may also play an important role in magnetic-resonance measurements in the presence of thermal gradients \cite{padronPRL11,*luPRL12,*jungfleischAPL13}. For thin ferromagnetic insulators, the interfaces could form a bottleneck for longitudinal spin transport with spin conductance $\sim(T/T_c)^{3/2}$. Our theory provides an essential building block for understanding the instabilities and dynamics of ferromagnets in the presence of thermal gradients in magnetic heterostructures or spin-transfer torque at finite temperature \cite{slonczewskiPRB10}. 

The authors thank S.~Takei and E.~G. Tveten for helpful discussions. This work was supported in part by the ARO under Contract No.~911NF-14-1-0016 and FAME (an SRC STARnet center sponsored by MARCO and DARPA).

\end{document}